\documentclass[11pt]{article}

\usepackage{comment}
\usepackage{graphicx}
\usepackage{url}
\usepackage{color}
\usepackage{natbib}
\usepackage{fullpage}
\usepackage{amsmath}
\usepackage{amssymb}
\usepackage{psfrag}
\usepackage{authblk}
\usepackage{multirow}
\usepackage{lineno}
\usepackage{setspace}
\usepackage{authblk}

\makeatletter
\let\start@align@nopar\start@align
\let\start@gather@nopar\start@gather
\let\start@multline@nopar\start@multline
\long\def\start@align{\par\start@align@nopar}
\long\def\start@gather{\par\start@gather@nopar}
\long\def\start@multline{\par\start@multline@nopar}
\makeatother

\newif\ifreview

\ifreview
\linenumbers
\fi

\renewcommand\footnotemark{}

\title{Residual cross-grid flow numerical error in large-eddy simulations of cumulus-topped boundary layers}

\author{Oumaima Lamaakel}
\author{Georgios Matheou}

\affil{Department of Mechanical Engineering, University of Connecticut, Storrs, Connecticut} 

\ifreview
\date{}
\else
\date{\today}
\fi

\begin{document}

\maketitle

\begin{abstract} 
A computational domain translation velocity is often used in LES simulations to improve  computational performance by allowing longer time-step intervals. Even though the equations of motion are Galilean invariant, LES results have been observed to depend on the translation velocity. It is found that LES results of shallow convection depend on the domain translation velocity even when a Galilean invariant formulation is used. This type of model error is named residual cross-grid flow error, to emphasize the expectation that it should be negligible or zero. The residual gross-grid flow error is caused by biases in finite difference dispersion errors. Schemes with low resolving power (typically low order of accuracy) produce larger dispersion errors that can be amplified by large-scale flow asymmetries, such as strong updrafts in cumulus-cloud layers. Accordingly, the cross-grid flow error strongly depends on the order of accuracy of the numerical scheme progressively becoming negligible as the order of accuracy is increased from second to sixth in the present simulations.
\end{abstract}

\section{Introduction}

Large-eddy simulation (LES) resolves all dynamically important flow scales and models the smaller, more ``generic'' in nature. In LES of atmospheric boundary layers resolutions typically range in 5--$50\; \rm m$ enabling simulations to explicitly resolve individual cloud shapes and the detailed structure of updrafts, downdrafts, and the entrainment process. The multiscale organization of convection requires large domains. In addition, long time integrations are needed to simulate the evolution of convection and diurnal cycle effects. 

Typically, time integrations with explicit time marching schemes are used, which adhere to an advection-dominated time-stability constraint. The time step length depends on the Courant--Friedrichs--Lewy (CFL) condition
\begin{equation}
\Delta t = \min_{ijk} \frac{\rm CFL_{max}}{ \frac{|u|}{\Delta x} + \frac{|v|}{\Delta y} + \frac{|w|}{\Delta z} }, \label{eq:cfl}
\end{equation}
where ${\rm CFL_{max}}$ is a scheme-dependent maximum CFL number, $[u, v, w]$ the velocity vector, $\Delta x$, $\Delta y$, and $\Delta z$ the spatial grid spacings, and the minimum is taken overall all grid cells. 

For a given computational domain size and simulation time length, the computation expense decreases as the grid size increases, i.e., $\Delta t$ is proportional to the spatial grid size $\Delta x$. 
Coarser grids are preferred, but because more spatial flow scales remain unresolved, LES models require skillful turbulence parameterizations to maintain the fidelity of the simulation as the grid becomes coarser \citep{Matheou_C.2014, Matheou_T.2019}. 

Another approach to gain computational advantage exploits the dependence of $\Delta t$ on the absolute value of the velocity field components. One one hand, $\Delta t$ increases for smaller absolute velocity components. On the other, the equations of motion are Galilean invariant and do not depend on the absolute value of the velocity. That is, the momentum equation is invariant under the transformation $\mathbf{u} \rightarrow \mathbf{u}+\mathbf{u}_0$ when $\mathbf{u}_0$ is a constant velocity vector. Accordingly, often in LES, the computation is performed in a moving frame, which is equivalent with a computational domain translating with a constant horizontal velocity with components $[u_0, v_0]$. The Galilean frame is chosen such that $|u_i|$ is minimized across all grid points, i.e., the computational domain translating with the domain-mean horizontal flow. Different domain translation velocities $\mathbf{u}_0$ result in changes in local velocity vector $\mathbf{u}(t,x,y,z)$ and, following the term of \cite{Wyant_BB.2018}, differences in cross-grid flow. 

In contrast to the continuous equations of motion, not all discrete approximations are Galilean invariant. The break down of Galilean invariance originates from non-linear discrete difference operators (i.e., approximations to the first derivative) that are typically used enforce the monotonicity property of the advected scalar fields. The error is introduced as artificial numerical dissipation that is proportional to a non-dimensional absolute velocity, such as a CFL number. 

The effects of variable numerical dissipation due to the frame of reference of the LES on the evolution of cloudy boundary layers have been investigated by \cite{Matheou_CNST.2011} and \cite{Wyant_BB.2018}. Even though the impact on boundary layer properties can be significant, and a violation of the Galilean invariance of the equations of motion, it is often a necessary tradeoff because non-linear numerical schemes are able to preserve the physical bounds of the scalar variables. Similar approaches exploiting moving grids to improve the overall accuracy of the simulations have been pursued in other disciplines as well \citep[e.g.,][]{Springel.2010}.

The dependence of LES on $\mathbf{u}_0$ when non-linear schemes are used is expected because the amount of dissipation is a function of $\mathbf{u}_0$ \citep{Matheou_CNST.2011, Wyant_BB.2018}. Presently, we investigate more subtle issues related to the Galilean invariance of LES when Galilean invariant numerical approximations are used. We demonstrate that even when linear finite difference schemes are used, results can depend on the choice of the frame of reference. We call this type of numerical model error as residual cross-grid flow error because it is not caused by parts of the derivative approximation truncation error that depend explicitly on $\mathbf{u}_0$. 

In linear finite difference methods the numerical derivative does not depend on the frame of reference, i.e., 
\begin{equation}
\left. \frac{{\rm  d} u}{{\rm d} x} \right|_{\rm fixed~frame} \approx  \sum_{p \ge i \ge q} \alpha_i u(x+i \, \Delta x) = \sum_{p \ge i \ge q} \alpha_i [ u(x+i \, \Delta x) + u_0] \approx \left. \frac{{\rm  d} u}{{\rm d} x} \right|_{u_0 \rm  ~frame},
\end{equation}
where $\sum_{p \ge i \ge q} \alpha_i = 0$ because of consistency. However, other sources of error can lead to departure from Galilean invariance. In LES, dispersion errors can be significant. As shown in \cite{Matheou_D.2016}, in a three-dimensional anisotropic turbulent flow, dispersion errors can preferentially occur upwind or downwind of the flow gradients depending on the numerical scheme used and the characteristics of the turbulence parameterization. 

Two causes of residual cross-grid model error are presently explored:~({\it a}) the presence of large-scale anisotropy, and ({\it b}) multi-phase flow effects that can potentially weaken some of the smoothness assumptions of the numerical approximation. 

Large-scale anisotropy typically occurs in regimes of cumulus convection where strong small-area updrafts rise in the conditionally unstable layer. In contrast, in ``dry'' (i.e., without water condensate) convection, updrafts and downdrafts have comparatively similar characteristics \citep{Chinita_MT.2018}. 

Because of the fine grid resolution in LES, the mean state of the grid cell is used to estimate the thermodynamic properties of the flow. That is, no subgrid variability is taken into account to classify each grid cell as saturated or clear. This leads to a spatially abrupt change in the thermodynamic coefficients. As shown in \cite{Grabowski_S.1990}, numerical artifacts can develop in an advection--condensation problem.  

Residual cross-grid model error is studied using simulations of dry and cumulus convection. Two types of simulations are carried out: ({\it a}) LES where the flow is not fully resolved and relatively large gradients are present at the grid scale, and ({\it b}) direct numerical simulation (DNS) where the flow is fully resolved and sufficiently smooth at the grid scale. In DNS, a turbulence closure is not used and the physical viscosity of the fluid provides all dissipation. Cloudy cases include simulations based on the conditions observed during the Cumulus over the Ocean (RICO) campaign \citep{Rauber_etal.2007.RICO, vanZanten_etal.2011} and buoyant bubble simulations, as a simple model for cumulus-topped updrafts. 

The model formulation and the simulation setups are presented in Section~2. The results are discussed in Section~3. Section~4 provided support for the large-scale anisotropy hypothesis . Finally, conclusions are summarized in Section~5.

\section{Methodology}
\subsection{Numerical model}
A unified numerical model is used to perform both large-eddy simulation and direct numerical simulation. When LES is carried out, a turbulence model is used to account for the effects of the unresolved motions on the resolved-scale variables and the contribution of the resolved scale viscous effects is neglected. When DNS is carried out, the turbulence model terms are not computed and all dissipation is provided by the viscous terms. 

The anelastic approximation of the conservation equations \citep{Ogura_P.1962} is numerically integrated on an $f$-plane (${\rm \{zonal, meridional, vertical\}}=\{x_1, x_2, x_3\}=\{x,y,z\}$). The conservation equations for mass, momentum, liquid water potential temperature, and total water, are, respectively,
\begin{equation}
\frac{\partial \bar{\rho}_0 \tilde{u}_i}{\partial x_i}=0, \label{eq:mass}
\end{equation}
\begin{equation}
\frac{\partial \bar{\rho}_0 \tilde{u}_i}{\partial t}+\frac{\partial (\bar{\rho}_0 \tilde{u}_i \tilde{u}_j)}{\partial x_j}=-\theta_0 \bar{\rho_0} \frac{\partial \bar{\pi}_2}{\partial x_i}+\delta_{i3} g \bar{\rho}_0 \frac{\tilde{\theta}_v - \langle \tilde{\theta}_v \rangle}{\theta_0}-\epsilon_{ijk}\bar{\rho}_0 f_j (\tilde{u}_k-u_{g,k})-\frac{\partial \tau_{ij}}{\partial x_j} + \frac{\partial  d_{ij}}{\partial x_j}, \label{eq:momentum}
\end{equation}
\begin{equation}
\frac{\partial \bar{\rho}_0 \tilde{\theta_l}}{\partial t} +\frac{\partial \bar{\rho}_0 \tilde{\theta_l} \tilde{u}_j}{\partial x_j}= -\frac{\partial \sigma_{\theta,j}}{\partial x_j} +\frac{\partial}{\partial x_j}\left( \rho_0 \mathcal{D}_\theta \frac{\partial \theta_l}{\partial x_j} \right) + \tilde{S}_\theta, \label{eq:theta}
\end{equation}
\begin{equation}
\frac{\partial \bar{\rho}_0 \tilde{q_t}}{\partial t} +\frac{\partial \bar{\rho}_0 \tilde{q_t} \tilde{u}_j}{\partial x_j}=-\frac{\partial \sigma_{q,j}}{\partial x_j}+ \frac{\partial}{\partial x_j}\left( \rho_0 \mathcal{D}_q \frac{\partial q_t}{\partial x_j} \right) + \tilde{S}_q. \label{eq:qt}
\end{equation}

The thermodynamic variables are decomposed into a constant potential temperature basic state, denoted by subscript 0, and a dynamic component. Accordingly, $\theta_0$ is the constant basic-state potential temperature and $\rho_0(z)$ is the density. The Cartesian components of the velocity vector and geostrophic wind, are $u_i$ and $u_{g,i}$, respectively and $f=[0,0,f_3]$ is the Coriolis parameter. Buoyancy is proportional to deviations of the virtual potential temperature $\theta_v$ from its instantaneous horizontal average $\langle \theta_v \rangle$, and $\pi_2$ is the dynamic part of the Exner function that satisfies the anelastic constraint (\ref{eq:mass}). 

When LES is performed, the prognostic variables $u_i$, $\theta_l$, and $q_t$ are defined as Favre-filtered variables $\tilde{\phi} \equiv \overline{\rho \phi}/\bar\rho$, where $\rho$ is the density and the overbar denotes a spatially filtered variable. When the flow is fully resolved (i.e., in DNS), $u_i$, $\theta_l$, and $q_t$ correspond to the local values (without any filtering or averaging), thus tildes and overbars are not needed. 

The viscous stress tensor is
\begin{equation}
d_{ij}=2 \mu D_{ij}
\end{equation}
where $\mu$ is the dynamic viscosity coefficient, which is assumed constant presently, and $D_{ij}$ is the deviatoric rate of strain tensor,
\begin{equation}
D_{ij}=\frac{1}{2}\left( \frac{\partial u_i}{\partial x_j} + \frac{\partial u_j}{\partial x_i} \right) - \frac{1}{3} \delta_{ij} \frac{ \partial u_k}{ \partial x_k}. \label{eq:dij}
\end{equation} 
The Fickian diffusion coefficients $\mathcal{D}_\theta$ and $\mathcal{D}_q$ are related to the momentum coefficient through the Prandtl and Schmidt numbers
\begin{equation}
Pr \equiv \frac{\nu}{\mathcal{D}_\theta} = 0.7,
\end{equation}
\begin{equation}
Sc \equiv \frac{\nu}{\mathcal{D}_q} = 1.
\end{equation}
where $\nu = \mu/ \rho_0$ is the kinematic viscosity. 

The subgrid-scale (SGS) stress tensor and scalar flux are modeled using an eddy-diffusivity assumption
\begin{equation}
\tau_{ij}=-2 \bar{\rho}_0 \nu_t \tilde{D}_{ij}, \label{eq:sgstensor}
\end{equation}
and
\begin{equation}
\sigma_{j,\phi}=-\bar{\rho}_0 \frac{\nu_t}{ \mathrm{Pr}_t} \frac{\partial \tilde{\phi}}{\partial x_j}.
\end{equation}
The eddy diffusivity for all scalar variables is related to the SGS momentum diffusivity, $\nu_t$, through the constant model turbulent Prandtl and Schmidt numbers, $\mathrm{Pr}_t=0.33$, $\mathrm{Sc}_t=0.33$.

The closure originally introduced by \cite{Smagorinsky.1963} and \cite{Lilly.1966, Lilly.1967} is used to estimate the turbulent diffusivity
\begin{equation}
\nu_t=\varDelta^2 | \tilde{D} | f_m(\mathrm{Ri}), \label{eq:nut}
\end{equation}
where $\varDelta=C_s \Delta x$ is the characteristics SGS length scale, $| \tilde{D} | = (2 \tilde{D}_{ij}\tilde{D}_{ij})^{1/2}$ is the resolved-scale deformation, and $f_m$ a stability correction function \citep{Lilly.1962},
\begin{align}
f_m &=
  \begin{cases}
   (1-\mathrm{Ri} /\mathrm{Pr}_t)^{1/2}         & \text{if }~\mathrm{Ri}/\mathrm{Pr}_t < 1 \\
      0        & \text{if } ~\mathrm{Ri}/\mathrm{Pr}_t \ge 1 \\
  \end{cases}
\end{align}
where $\mathrm{Ri}=N^2/|\tilde{D}|^2$ is the gradient Richardson number and $N$ is the buoyancy frequency. The value $C_s = 0.2$ is used for the Smagorinsky constant based on the parametric study of \cite{Matheou.2016}. The constant-coefficient Smagorinsky turbulence model is used because it can be directly observed from (\ref{eq:dij}) and (\ref{eq:nut}) that it only depends on derivatives of the flow fields. Galilean invariance is a necessary property of any turbulence model. 

Near the surface, the characteristic length scale is modified to account for the confinement of the SGS eddies \citep{Mason_C.1986}, 
\begin{equation}
\frac{1}{\varDelta^2}=\frac{1}{(C_s \Delta x)^2}+\frac{1}{(\kappa z)^2},
\end{equation}
where $\kappa=0.4$ is the von K\'{a}rm\'{a}n constant and $z$ the height from the surface.

The effect of the large-scale environment and clear air radiative cooling is included in the equations for $\theta_l$ and $q_t$ through the source terms $S_\theta$ and $S_q$.  

Condensation is modeled based on the mean thermodynamic state in each grid cell. For all but one pair of simulations an ``all or nothing'' scheme is used, i.e., no partially saturated air in each grid cell is assumed. Two runs use a modified saturation scheme that allows for the presence of liquid when the mean state is not saturated. 

The liquid water mixing ratio $q_l$ in the ``all or nothing'' scheme is
\begin{equation}
q_l = \max(0, q_t - q_s), \label{eq:saturation}
\end{equation}
where $q_s(p, T)$ is the saturation mixing ratio. Equation (\ref{eq:saturation}) is the typical saturation model used in LES of atmospheric boundary layers. 

An ad hoc subgrid condensation scheme is constructed by using a smoother transition between the unsaturated and saturated regimes
\begin{align}
q_l &=
  \begin{cases}
   0        & \text{if }~~~~ q_t - q_s < -0.5 \rm \; g\,kg^{-1} \\
      500(q_t - q_s)^2+0.5(q_t - q_s)+0.000125        & \text{if } ~~~~ -0.5 \le q_t - q_s \le 0.5 \rm \; g\,kg^{-1} \\
  q_t - q_s       & \text{if}~~~~ q_t - q_s > 0.5 \rm \; g\,kg^{-1}. \label{eq:mod_saturation}
  \end{cases}
\end{align}
In the modified saturation scheme (\ref{eq:mod_saturation}) a second degree polynomial is used to transition between the two branches of (\ref{eq:saturation}).

Liquid water is assumed suspended (i.e., no drizzle or precipitation is present) in all simulations, even though for the shallow cumulus case precipitation develops as the boundary layer deepens \citep{vanZanten_etal.2011}. 

Spatial derivatives are approximated with centered finite difference approximations. The family of fully conservative schemes of \cite{Morinishi_LVM.1998}, adapted for the anelastic approximation, is used for the momentum and scalar advection terms. The second-, fourth-, and sixth-order approximations are used. The properties of the advection schemes are discussed in \cite{Matheou.2016} and \cite{Matheou_D.2016}. The key difference between the three approximations is the resolving power: the property of faithfully representing the finer-scale motions for a given grid resolution, which is more important in LES than the formal order of accuracy \citep[e.g.,][]{Hill_P.2004}. For the present schemes the resolving power increases with increasing order of accuracy. For all cases, regardless of the order of the advection scheme, second-order centered differences are used to approximate the spatial derivatives of the viscous and subgrid scale model terms. The semi-discrete system of equations is advanced in time using the third-order Runge--Kutta of \cite{Spalart_MR.1991}. 

All simulations are preformed in a doubly-periodic domain in the horizontal directions. A Rayleigh damping layer is used at the top of the domain to limit gravity wave reflection. 

\subsection{Simulations}

\subsubsection{Common forcing}

To create similar wind profiles, all simulations use the geostrophic wind forcing of the LES case of the Cumulus over the Ocean (RICO) field study \citep{vanZanten_etal.2011}. The components of the geostrophic wind are $u_g (z) = -9.9 + 0.5\times 10^3 z \; \rm m\,s^{-1}$ and $v_g = -3.8 \; \rm m\,s^{-1}$. The latitude is $18^\circ \; \rm N$. The differences between the present cases are created mainly by different initial temperature and humidity profiles and surface fluxes. All cases are run in pairs, with and without a Galilean translation velocity $\mathbf{u}_0$. The translation velocity for the dry convection and shallow cumulus cases is $(-6, -4) \; \rm m\,s^{-1}$. For the buoyant bubble cases the translation velocity is $(-9, -3.8) \; \rm m\,s^{-1}$. The difference is because the buoyant bubble case does not include surface shear, thus the mean wind is somewhat different. 

For all cases the grid resolution is typical of similar studies in the literature. Grid spacing is uniform and isotropic $\Delta x = \Delta y = \Delta z$. The time step is adjusted to maintain $\rm CFL = 1.2$. The CFL numerical stability limit is $\rm CFL_{max} = \sqrt{3}$. Table~\ref{table:runs} summarizes the LES runs. 

\subsubsection{Dry convection}

The cloud-free (i.e., ``dry'') convection case of \cite{Matheou_CNST.2011} is modified by the addition of geostrophic wind forcing. The initial potential temperature lapse rate is $2 \; \rm K\,km^{-1}$, with $\theta(z=0) = 297 \; \rm K$. The initial total water mixing ratio lapse rate is $-0.37 \; \rm g  \, kg^{-1} \, km^{-1}$ up to $z = 1350\; \rm m$  and $-0.94 \; \rm g  \, kg^{-1} \, km^{-1}$ higher up with $q_t(z=0) = 5 \; \rm g \, kg^{-1}$. The temperature and humidity surface fluxes are $0.06 \; \rm K\, m\,s^{-1}$ and $2.5 \times 10^{-5} \; \rm m\,s^{-1}$, respectively. The surface shear stresses are computed in each grid cell using the Monin--Obukhov similarity theory. The simulations are run for $4 \; \rm h$. 

\subsubsection{Buoyant bubble}

The temperature and humidity initial profiles of the RICO case are used. An initial spherical positively buoyant region with radius  $r_0 = 200 \; \rm m$ and center at $z=r_0$ is created by increasing the values of $\theta_l$ and $q_t$ by $10\,\%$ with respect to the standard (horizontally uniform) initial condition. The initial condition is given by
\begin{equation}
\phi(x, y, z)=[0.05 \, {\rm erf}(0.05(r_0-r))+1.05] \, \phi_i(z)
\end{equation}
where $r=\left(x^2+y^2+(z-r_0)^2\right)^{1/2}$ is the distance from the center of the sphere, $\phi$ denotes either $\theta_l$ or $q_t$, and $\phi_i(z)$ the initial profile of the RICO case. 

The large-scale forcing of the RICO case is not included in the buoyant bubble simulations. Sensible and latent heat surface fluxes are set to zero.  

LES and DNS types of simulations are carried out. Two pairs of LES at grid resolutions $\Delta x = 10 \; \rm m$ and $\Delta x = 20 \; \rm m$ are performed. 

In the DNS run, viscosity is set to $\mu = 2 \; \rm kg \, m^{-1} \, s^{-1}$, which results in smooth well-resolved fields for $\Delta x = 5 \; \rm m$. The turbulence model terms are set to zero in the DNS runs. Because surface shear cannot be resolved, a slip (no penetration, no stress) surface condition is used. The DNS simulations are run for $1 \; \rm h$ and the LES simulations for $0.5 \; \rm h$. 

\subsubsection{Shallow cumulus convection}

The shallow cumulus convection simulations follow the setup of the RICO case but do not include the process of precipitation. The RICO conditions are chosen because convection is more vigorous compared to other cases of non-precipitating shallow convection, e.g., \cite{Siebesma_etal.2003}, therefore, it is expected to be a more stringent case. The initial $\theta$ and $q_t$ profiles have a mixed layer depth of $740 \; \rm m$ and linearly decrease above the mixed layer. Large scale subsidence moisture and humidity advection and a uniform clear sky radiative cooling are included in the simulations. The surface fluxes are parameterized using bulk transfer coefficients and a constant sea surface temperature $298.8 \; \rm K$. Details of the case setup are described in \cite{vanZanten_etal.2011}. The simulations are run for $18 \; \rm h$.

\begin{table}[t]
\caption{Summary of the cases simulated. The first and second columns correspond to the shortened form of the simulation case and the convection type, respectively. Fully resolved simulations, without any turbulence parameterization, are denoted as DNS, whereas simulations of the full dynamics using a turbulence closer are labeled as LES. The grid spacing is denoted by $\Delta x$, $N_x= N_y$ and $N_z$ are number of horizontal and vertical grid points, respectively, $\mathbf{u}_0$ the the Galilean translation velocity, and ``Advection'' corresponds to the order of the advection scheme. For all runs $\Delta x = \Delta y = \Delta z$. The star ($^*$) in case Df denotes that a 10-member ensemble was carried out.}\label{table:runs}
\begin{center}
\begin{tabular}{llcccccccccc}
\hline\hline
Run & Description & Model & $\Delta x$  &  $N_x$ & $N_z$ & $\mathbf{u}_0$  & Advection\\
\noalign{\smallskip}\hline\noalign{\smallskip}
BDf & Buoyant bubble & DNS & 5 & 512 & 600 & $(0,0)$ & fourth \\
BDg & Buoyant bubble & DNS & 5 & 512 & 600 & $(-9, -3.8)$ & fourth \\

BHf & Buoyant bubble & LES & 10 & 256 & 300 & $(0,0)$ & fourth \\
BHg & Buoyant bubble & LES&  10 & 256 & 300 & $(-9, -3.8)$ & fourth \\
BLf & Buoyant bubble & LES&  20 & 256 & 300 & $(0, 0)$ & fourth \\
BLg & Buoyant bubble & LES&  20 & 256 & 300 & $(-9, -3.8)$ & fourth \\

C2f & Shallow Cu & LES & 40 & 1024 & 100 &   $(0,0)$ & second \\
C2g & Shallow Cu & LES & 40  & 1024 & 100 & $(-6,-4)$ & second \\
C4f & Shallow Cu & LES & 40 & 1024 & 100 &   $(0,0)$ & fourth \\
C4g & Shallow Cu & LES & 40  & 1024 & 100 & $(-6,-4)$ & fourth \\
C6f & Shallow Cu & LES & 40 & 1024 & 100 &  $(0,0)$ & sixth \\
C6g & Shallow Cu & LES & 40  & 1024 & 100 &  $(-6,-4)$ & sixth \\
CSf & Shallow Cu (mod. saturation) & LES & 40  & 1024 & 100  & $(0,0)$ & fourth\\
CSg & Shallow Cu (mod. saturation) & LES & 40  & 1024 & 100  & $(-6, -4)$ & fourth\\
Df$^*$ & Dry convection & LES & 40 & 512 & 100 & $(0,0)$ &  fourth \\
Dg & Dry convection & LES & 40 & 512 & 100 & $(-6,-4)$ & fourth \\
\end{tabular}   
\end{center}
\end{table}

\section{Results}

\subsection{Dry convection}

\begin{figure}[t!]
\centering
\includegraphics[width=\columnwidth]{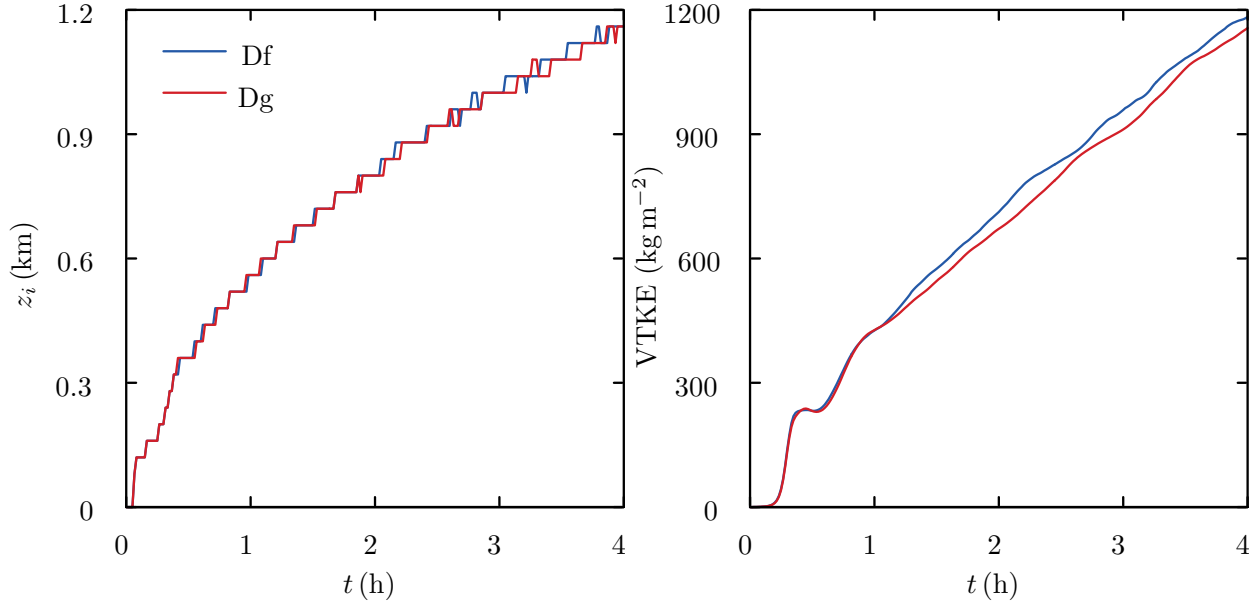} 
\caption{Time evolution of the boundary layer height $z_i$ and vertically integrated turbulent kinetic energy for a dry convective boundary layers in the fixed (Df) and Galilean (Dg) frames.} \label{fig:dcbl_traces}
\end{figure}

\begin{figure}[t!]
\centering
\includegraphics[width=\columnwidth]{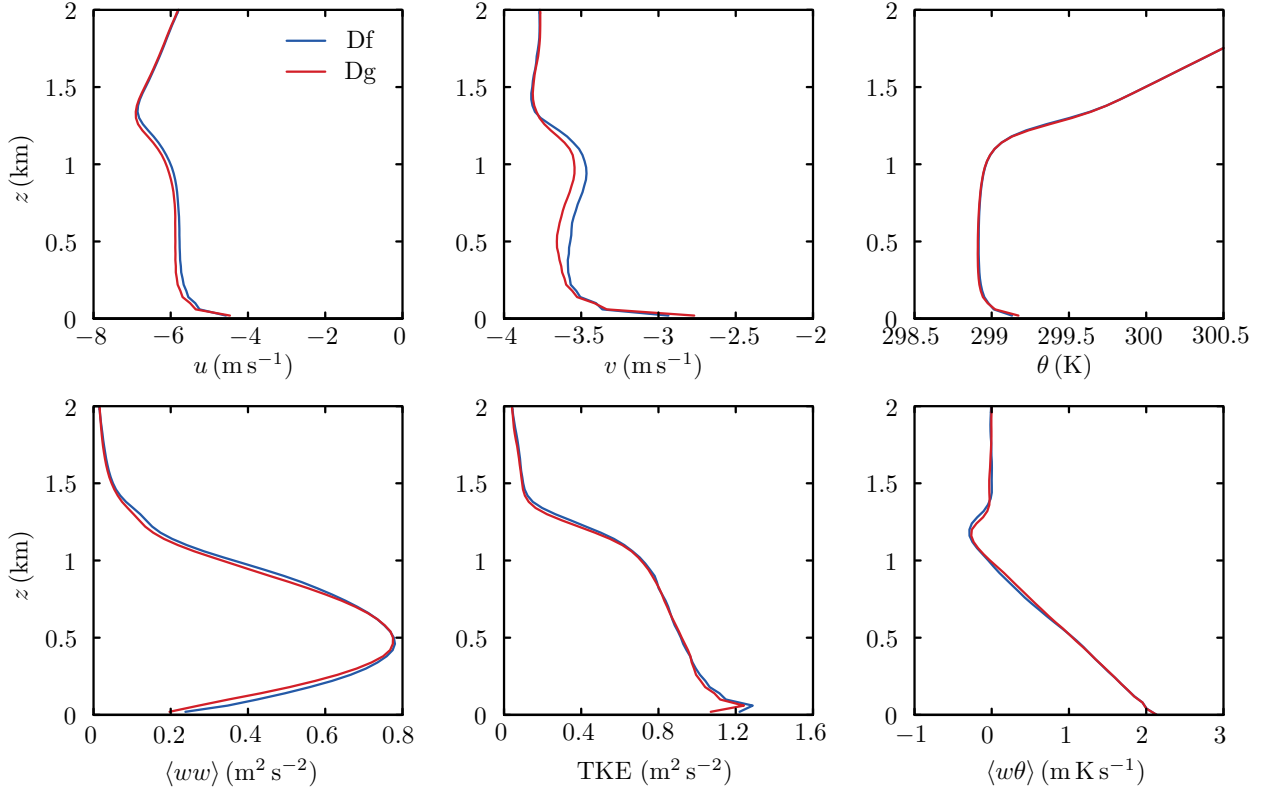} 
\caption{Dry convective boundary layer profiles of zonal $u$ and meridional wind $v$, potential temperature $\theta$, vertical velocity variance $\langle ww \rangle$, resolved scale turbulent kinetic energy, and temperature flux $\langle w \theta \rangle$ at $t= 4 \; \rm h$ (not time averaging) for simulations in the fixed (Df) and Galilean (Dg) frames. The turbulent fluxes are the sum of the resolved scale and subgrid scale components.} \label{fig:dcbl_profiles}
\end{figure}

The effects of domain translation velocity are negligible for the dry convection case. Figure~\ref{fig:dcbl_traces} shows time traces of boundary layer height $z_i$, defined as the height of the minimum of the buoyancy flux, and vertically integrated turbulent kinetic energy (VTKE). Figure~\ref{fig:dcbl_profiles} shows profiles at the end of the run $t = 4 \; \rm h$. 

In all cases, the profiles are instantaneous horizontal averages, i.e., no time averaging is performed. Also, scalar turbulent fluxes $\langle w \theta \rangle$ and $\langle w q_t \rangle$ include the subgrid scale contribution. The isotropic part of $\tau_{ij}$ is not explicitly available in the Smagorinsky closure, it is part of the dynamic pressure. Thus, TKE profiles correspond only to the resolved scale. 

The small differences between the runs of Figs.~\ref{fig:dcbl_traces} and \ref{fig:dcbl_profiles} are more likely because of statistical variability rather than cross-grid flow errors. The range of statistical variability of VTKE (see Appendix A) is comparable to the differences between the VTKE traces (fig.~\ref{fig:dcbl_traces}), thus the differences between the fixed and Galilean frames are not statistically significant. Moreover, VTKE amplifies the differences in the TKE profiles because it is an integral measure through the height of the boundary layer. 

\subsection{Shallow cumulus}

\begin{figure}[t!]
\centering
\includegraphics[width=\columnwidth]{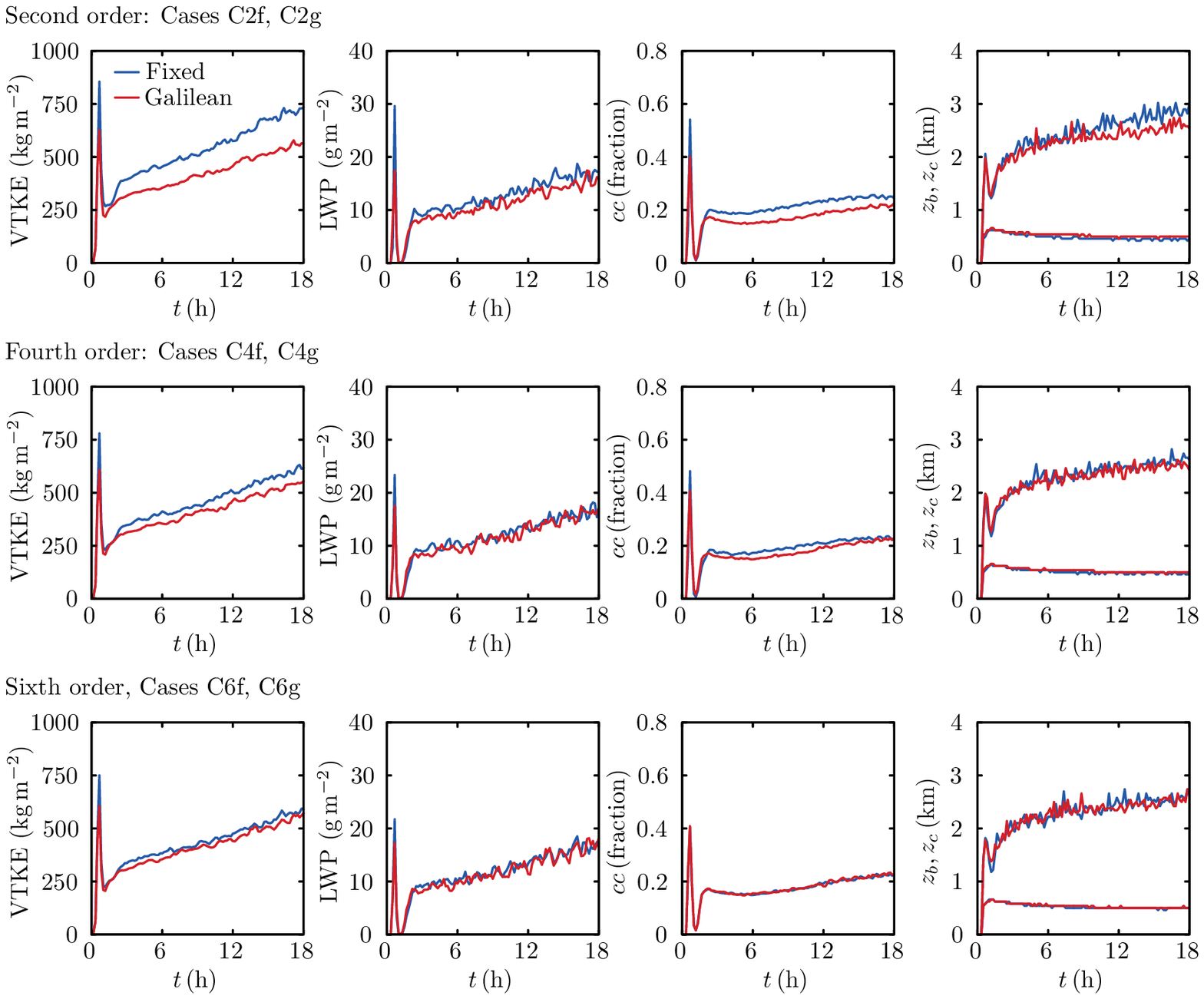} 
\caption{Time evolution of the vertically integrated turbulent kinetic energy (VTKE), liquid water path (LWP) cloud cover, and cloud base and top heights for the shallow cumulus cases in the fixed (blue lines) and Galilean (red lines) frames. Each row of panels corresponds to different advection scheme order of accuracy.} \label{fig:rico_traces}
\end{figure}

Shallow cumulus results show dependence on $\mathbf{u}_0$ with differences depending on the advection scheme order. The differences of VTKE, LWP, cloud cover, cloud base, and cloud top traces (fig.~\ref{fig:rico_traces}) are larger when the second order scheme is used and negligible for the sixth-order scheme. Moreover, the sensitivity depends on the flow statistic: LWP and the boundary layer depth are less sensitive to the frame of reference and only simulations using the second-order scheme show differences. VTKE and cloud cover are the most sensitive quantities. Cloud cover $cc$ is defined as the fraction of columns with at least one level where $q_l > 10^{-5} \; \rm kg\,kg^{-1}$. As defined, cloud cover is very sensitive to horizontal fluctuations of small $q_l$ values. LWP is more representative of the cumulus characteristics. Only simulations with the second order scheme show (small) differences in LWP with respect to $\mathbf{u}_0$.  

\begin{figure}[t!]
\centering
\includegraphics[width=\columnwidth]{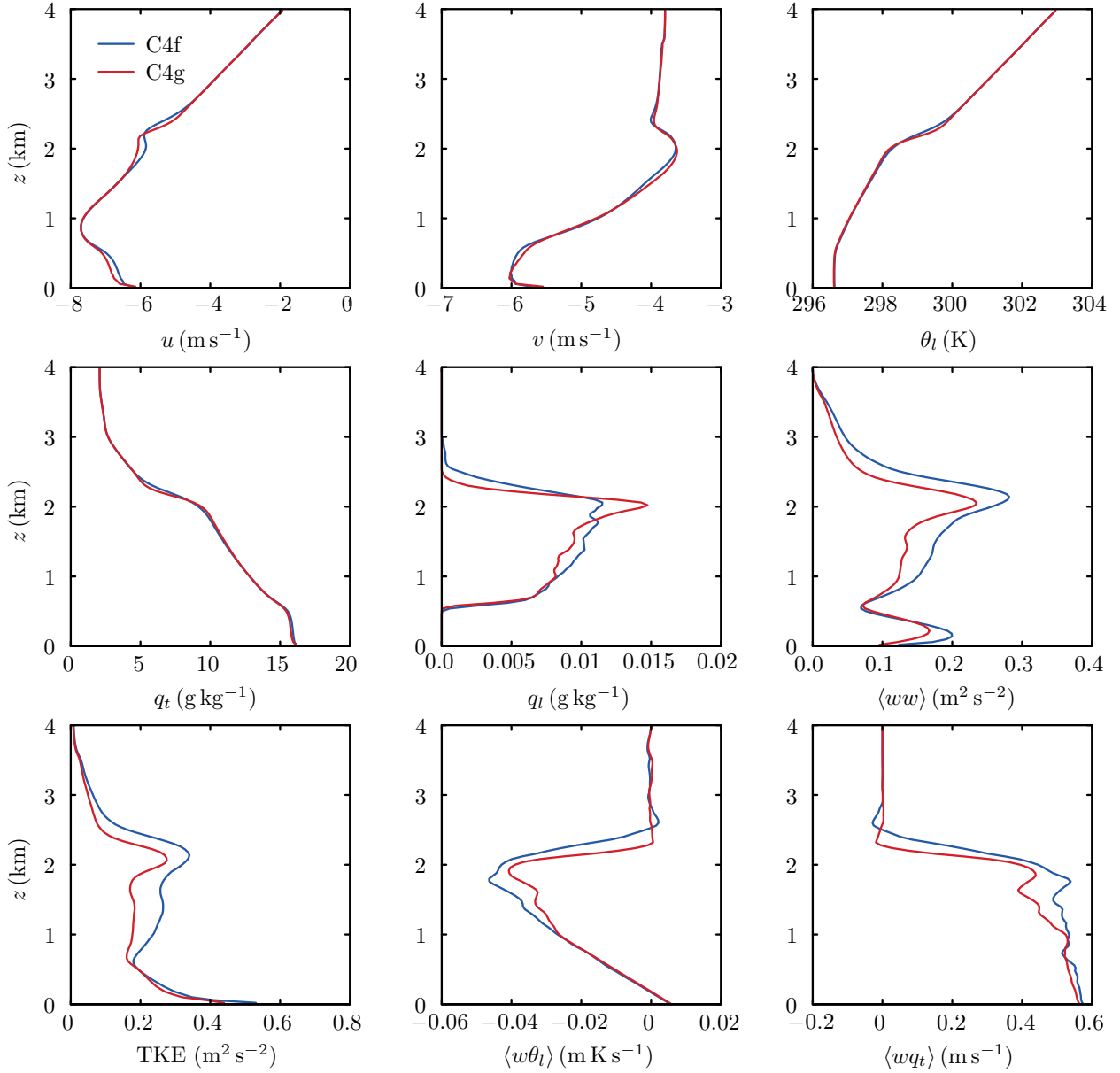} 
\caption{Shallow cumulus profiles (cases C4f and C4g) of zonal $u$ and meridional wind $v$, liquid water potential temperature $\theta_l$, total water mixing ratio $q_t$, liquid water mixing ratio $q_l$, vertical velocity variance $\langle ww \rangle$, resolved scale turbulent kinetic energy, temperature flux $\langle w \theta_l \rangle$, and total water flux $\langle w q_t \rangle$ at $t= 18 \; \rm h$ (not time averaging) for simulations in the fixed (C4f) and Galilean (C4g) frames. The turbulent fluxes are the sum of the resolved scale and subgrid scale components.} \label{fig:rico_profiles}
\end{figure}

A more detailed view of the differences for the pair of simulations using the fourth-order scheme is shown in the profiles of fig.~\ref{fig:rico_profiles}. Differences in the mean fields of the prognostic variables are negligible but turbulent fluxes and $q_l$ differ, particularly TKE and $\langle ww \rangle$. The differences are mostly in the cloud layer. Only $\langle ww \rangle$ is different in the lower half of the mixed layer. This result is consistent with the dry convection and suggests that the discrepancies with respect to $\mathbf{u}_0$ are triggered in the cloud layer.

In Figure~\ref{fig:rico_traces}, the Galilean frame results do not change with respect to the advection scheme and the fixed-frame LES converges towards the Galilean frame traces as the order of accuracy in increased. Thus, very likely the Galilean frame corresponds to the LES with the least cross-grid flow error. The amount of VTKE error for the second-order scheme is somewhat surprising, given that only the order of accuracy of the momentum and scalar advection schemes changes between C2f and C6f cases. Similar  advection discretization effects have been observed in LES of stable boundary layers \citep{Matheou.2016}.

\begin{figure}[t!]
\centering
\includegraphics[width=\columnwidth]{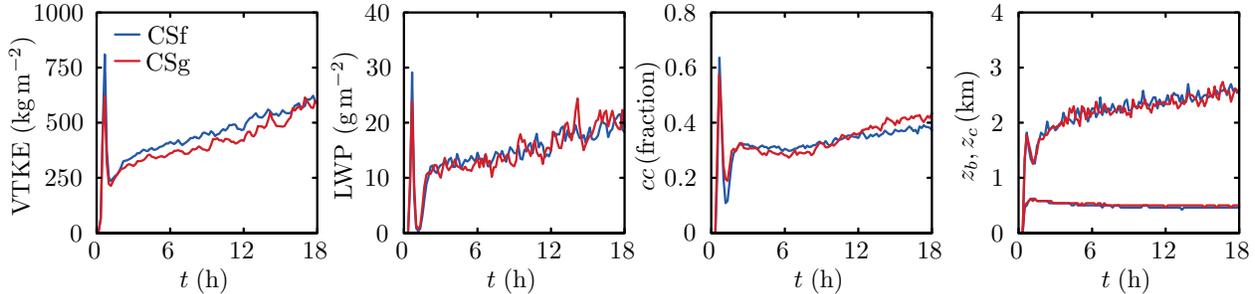} 
\caption{Time evolution of the vertically integrated turbulent kinetic energy (VTKE), liquid water path (LWP) cloud cover, and cloud base and top heights for the shallow cumulus simulations with the modified saturation scheme in the fixed (CSf) and Galilean (CSg) frames.} \label{fig:rico_cf_traces}
\end{figure}

A second pair of simulations using a modified condensation scheme [Eq. (\ref{eq:mod_saturation})] was carried out to assess if the differences are because of condensation/evaporation effects. Figure~\ref{fig:rico_cf_traces} shows time traces for the pair of LES with the modified condensation scheme. The differences in VKTE with respect to the frame of reference are the same as in the corresponding runs using the ``all or nothing'' condensation scheme, C4f and C4g. LWP and cloud cover increase in runs CSf/g since additional partial condensation occurs in the modified scheme.

\subsection{Buoyant bubble}

\begin{figure}[t!]
\centering
\includegraphics[width=\columnwidth]{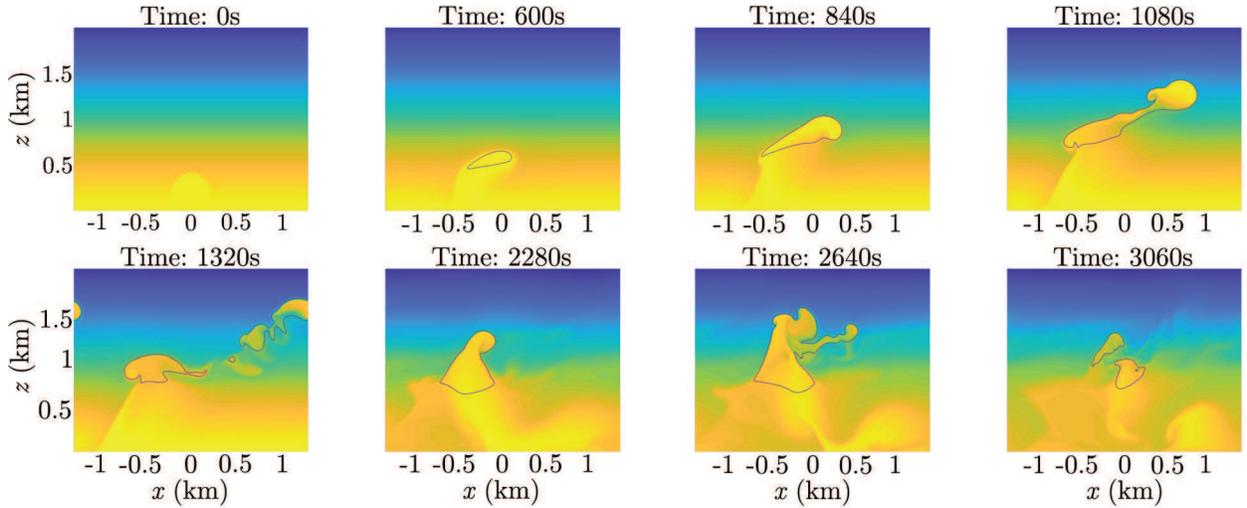} 
\caption{Direct numerical simulation of a buoyant bubble. Color contours show the evolution of total water mixing ratio. Black contour corresponds to the saturation mixing ratio, denoting the cloud boundary.} \label{fig:bubble}
\end{figure}

\begin{figure}[t!]
\centering
\includegraphics[width=\columnwidth]{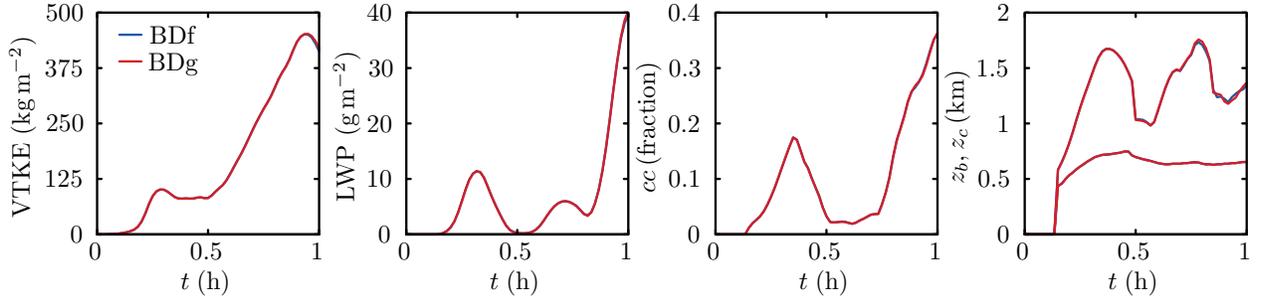} 
\caption{Comparison of the time evolution of vertically integrated turbulent kinetic energy, LWP, cloud cover $cc$, and cloud base $z_b$ and height $z_c$ for the buoyant bubble direct numerical simulations in the fixed (blue lines) and Galilean frames.} \label{fig:bubble_laminar}
\end{figure}

The buoyant bubble simulations are a simplified model of convection. The simpler configuration allows simulation of a pair of cases where the flow is fully resolved using constant viscosity coefficients, thus creating an effective DNS. Figure~\ref{fig:bubble} shows the evolution of the flow in the DNS case. The flow is initially driven by potential energy. The rising bubble in a flow with mean shear creates a fairly complex flow. At about $t = 1200 \; \rm s$ the initial bubble reaches a top height $z \approx 1.5 \; \rm km$ (fig.~\ref{fig:bubble_laminar}). The disturbance caused by the bubble rise and entrainment results in some of the near-surface air to reach the level of free convection, thus a secondary cloud-top plume is created after $t = 1800 \; \rm s$ (fig.~\ref{fig:bubble}). 

\begin{figure}[t!]
\centering
\includegraphics[width=\columnwidth]{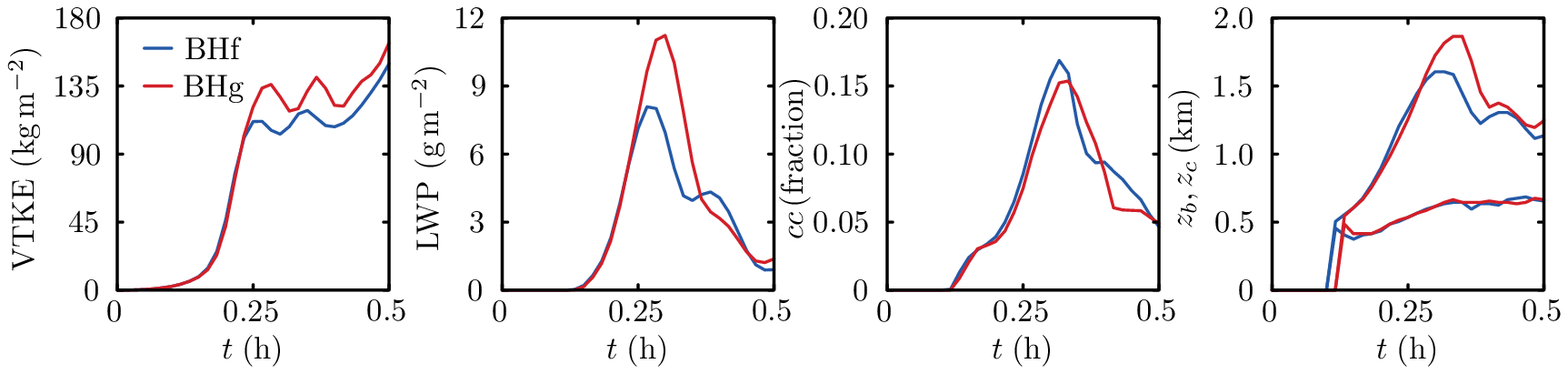} 
\caption{Time evolution of the vertically integrated turbulent kinetic energy (VTKE), liquid water path (LWP) cloud cover, and cloud base and top heights for the high resolution ($\Delta x = 10 \; \rm m$) buoyant bubble simulations in the fixed (BHf) and Galilean (BHg) frames.} \label{fig:bubble_dx10}
\end{figure}

\begin{figure}[t!]
\centering
\includegraphics[width=\columnwidth]{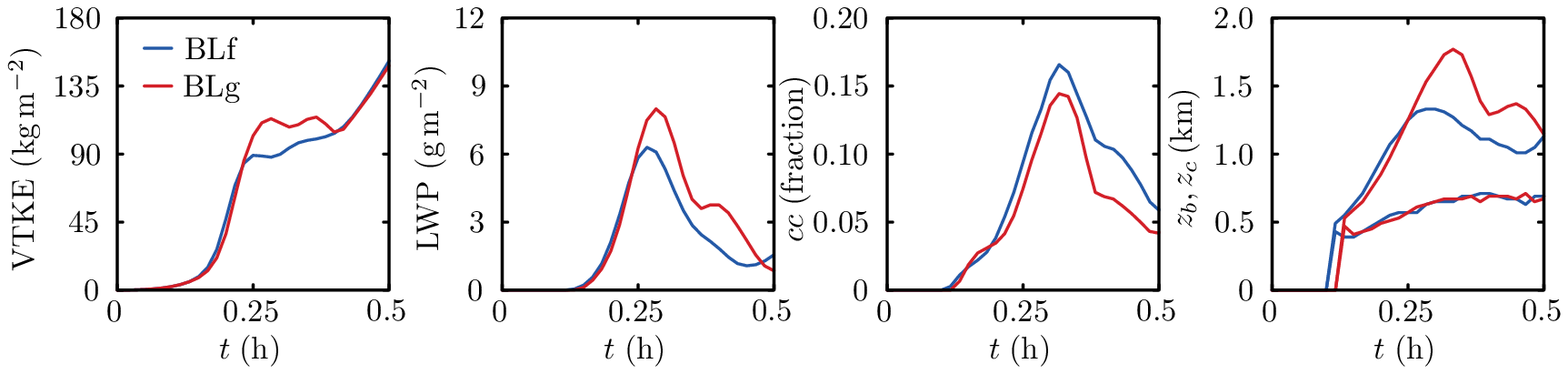} 
\caption{Time evolution of the vertically integrated turbulent kinetic energy (VTKE), liquid water path (LWP) cloud cover, and cloud base and top heights for the low resolution ($\Delta x = 20 \; \rm m$) buoyant bubble simulations in the fixed (BLf) and Galilean (BLg) frames.} \label{fig:bubble_dx20}
\end{figure}

Comparison of the LES runs at $\Delta x = 10$ and $20 \; \rm$ are shown in figs.~\ref{fig:bubble_dx10} and \ref{fig:bubble_dx20}, respectively. Both grid resolutions show similar differences with respect to $\mathbf{u}_0$. The differences appear only after the flow developed rich three-dimensional structure $t > 0.5 \; \rm h$.   In the shallow cumulus simulations the cases in the fixed frame had larger VTKE and LWP. However, the buoyant bubble simulations show the opposite trend, suggesting that the differences can be of either sign. It is likely that the difference sign depends on the flow geometry. 

The results of the fully resolved simulation (fig.~\ref{fig:bubble_laminar}) are identical in the two frames of reference and confirm that the numerical discretization is Galilean invariant.

\section{Discussion}

The present LES results show that the differences in the fixed and Galilean frames can occur in cloudy convection. The error is likely triggered by the structure of the flow, rather than variations of the buoyancy forcing because condensation and evaporation. The fully resolved simulations confirm that numerical method is Galilean invariant. 

\begin{figure}[t!]
\centering
\includegraphics[width=\columnwidth]{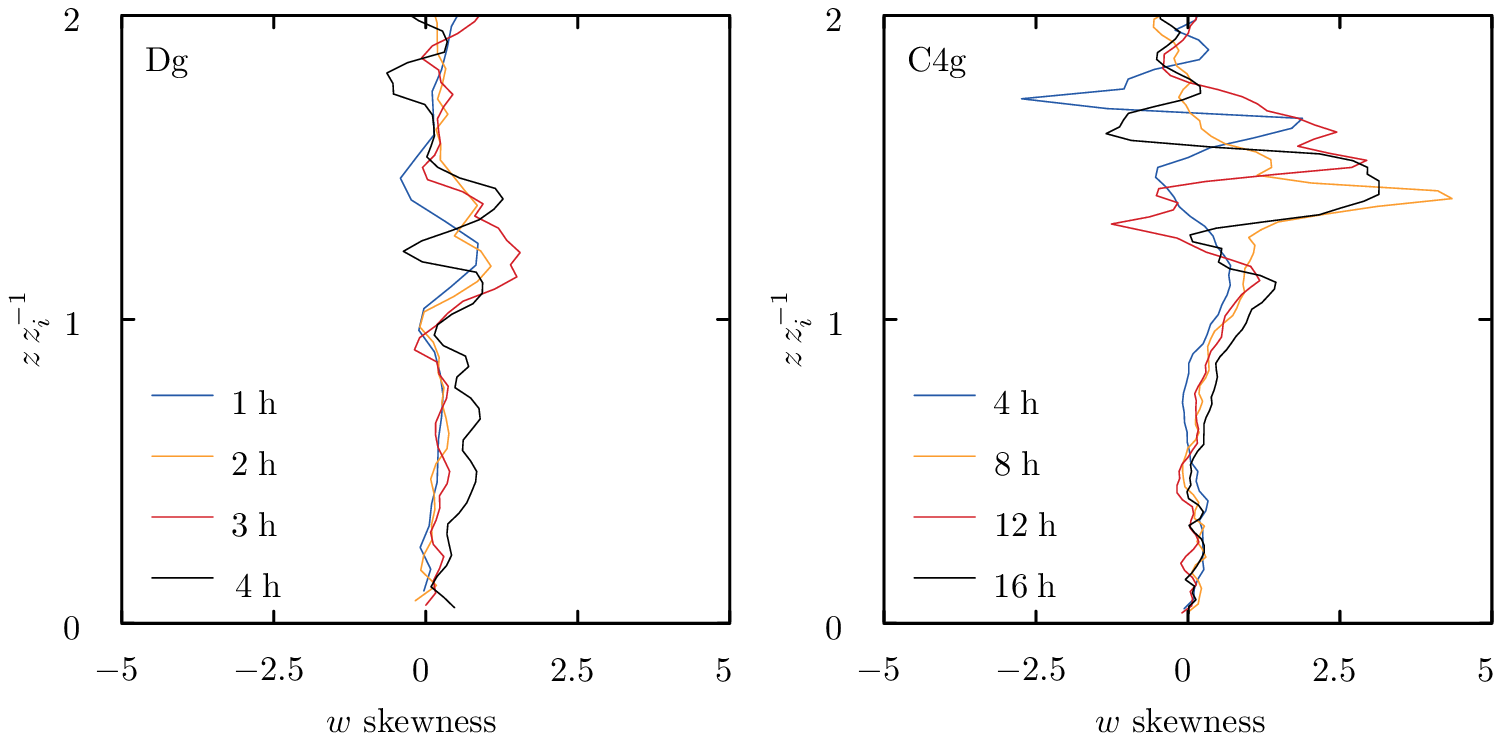} 
\caption{Vertical velocity skewness profiles at different times for the dry convection (Dg, top) and shallow cumulus (C4g) cases. The vertical axis is scaled by the height of the minimum buoyancy flux $z_i$.} \label{fig:skewness}
\end{figure}

The differences in the flow structure are quantified by considering the skewness of the vertical velocity. Figure~\ref{fig:skewness} shows $w$ skewness for a dry, Dg, and a cumulus case, C4g, at four instances during the run. Height is normalized with $z_i$, the height of the minimum buoyancy flux. The normalization with $z_i$ scales $z$ with the depth of the mixed layer. As a consequence, for the dry convection case turbulence is confined in the layer $< 1.2 z / z_i$, whereas in the shallow cumulus case, the boundary layer grows in time to reach about $2 z/ z_i$ by the end of the run. As shown in the cumulus case, $z_i$ scales the $w$ skewness well for $z / z_i < 1$. 

As expected \citep{Heus_J.2008}, the vertical velocity distribution is positively skewed in the cloud layer because of the strong updrafts in the cloud cores. In the subcloud layer and in dry convection cases, the positive bias of the $w$ distribution is significantly less creating a flow with more symmetric structure. 

\section{Conclusions}

A computational domain translation velocity $\mathbf{u}_0$ is often used in LES simulations to improve  computational performance by allowing larger time steps. Even though the equations of motion are Galilean invariant, i.e., do not depend on $\mathbf{u}_0$, LES results have been observed to depend on $\mathbf{u}_0$ \citep{Matheou_CNST.2011, Wyant_BB.2018}. Non-linear numerical schemes that are not Galilean invariant can result in differences in the LES results with respect to $\mathbf{u}_0$, because artificial numerical dissipation depends on $\mathbf{u}_0$. 

We show that it is possible for LES results of shallow convection to depend on $\mathbf{u}_0$ even when a Galilean invariant formulation is used. Because this form of model error should not occur based on the scheme formulation properties, we call this a residual cross-grid flow error, to emphasize the expectation that it should be negligible or zero. 

Residual cross-grid flow errors mostly affect second-order statistics, including liquid water profiles and liquid water path, and, to a lesser extend, boundary layer growth rates. In the present simulations, the most sensitive quantity is the turbulent kinetic energy (TKE). The vertical integral of TKE was found to differ by as much as $20 \%$ between fixed-frame and moving frame LES. The sign of the difference likely depends on the flow structure. 

The present results suggest that the residual gross-grid flow error is caused by biases in finite difference dispersion errors. Low order, or more accurately, schemes with less resolving power, produce larger dispersion errors that can be amplified by large-scale flow asymmetries, such as strong updrafts in cumulus-cloud layers. Accordingly, the cross-grid flow error strongly depends on the order of accuracy of the numerical scheme progressively becoming negligible as the order is increased from second to sixth. 
A pair of fully resolved direct numerical simulations (DNS), which have negligible dispersion errors, verify the Galilean invariance of the method and confirm that gross-grid errors vanish for smooth flow fields. 

Residual cross-grid errors are not an artifact of the advection--condensation/evaporation problem, even though the error is negligible in non-cloudy convection cases. The comparison of vertical velocity skewness between the cumulus and non-cloudy convection cases supports the argument flow anisotropy facilitates the growth of residual cross-grid flow errors. 

\subsection*{Acknowledgements}
The research presented in this paper was supported by the systems, services, and capabilities provided by the University of Connecticut High Performance Computing (HPC) facility. 

\appendix
\section{Statistical variability of turbulent kinetic energy}
\begin{figure}[t!]
\centering
\includegraphics[width=7cm]{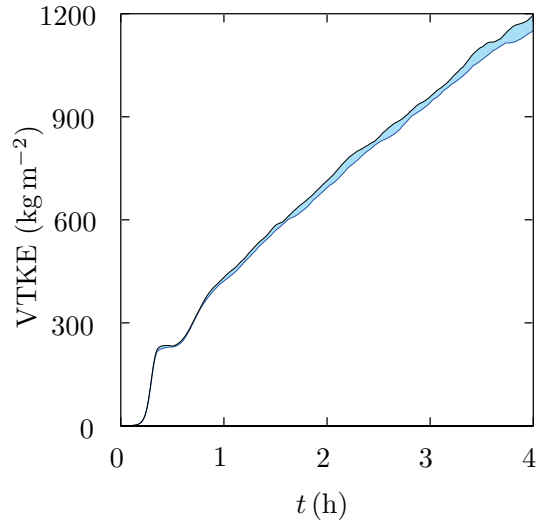} 
\caption{Spread of vertically integrated turbulent kinetic energy vs time for the ten-member-ensemble dry convective boundary layer.} \label{fig:dcbl_tke_range}
\end{figure}

A ten-member ensemble is carried out to estimate the statistical variability of VTKE in the dry convection case. Because of the finite computational domain, a complete sample of the flow states is not accomplished and instantaneous horizontal averages are not fully converged. Figure~\ref{fig:dcbl_tke_range} show the band of VTKE variability of the ensemble. As the boundary layer deepens, the convection cells become larger and fewer in the fixed domain size. Thus, the sampling of the flow declines with time and the band of VTKE variability widens with respect to time. After $t = 3 \; \rm h$ VTKE is uncertain by about $50 \; \rm kg\,m^{-2}$ or $4\%$.

\bibliography{bibliography}
\bibliographystyle{./ametsoc}

\end{document}